# EFFICIENCY IMPROVEMENT OF COMMERCIALLY AVAILABLE MPPT CONTROLLERS USING BOOST CONVERTER

**Lean Karlo S. Tolentino**
**Dan Frederico P. Abacco**
**Mar Jun M. Siquihod**

Department of Electronics Engineering
Technological University of the Philippines
Ayala Blvd., Ermita Manila, Philippines
leankarlo_tolentino@tup.edu.ph, danfrederico.abacco@tup.edu.ph, marjun.siquihod@tup.edu.ph

## ABSTRACT

The use of charge controllers in photovoltaic generation systems increases the energy harvested significantly. The most efficient type of charge controller is the Maximum Power Point Tracking (MPPT) controller. The commercially available MPPT controller is limited to the input voltage of the solar panel and to the charging voltage of the battery making it inefficient to very low irradiance weather condition. In this paper, it is proposed to increase the efficiency of commercially available MPPT controllers by adding Power Management System that has boost converter and switching circuit that will operate at low irradiance weather conditions like cloudy, sunrise, and sunset. The boost converter is controlled by the PWM function of Arduino Uno to regulate the voltage required of the MPPT controller. The switching circuit will automatically change the direction of current depending on the output of the voltage sensors attached to the PV panel and input of MPPT controller. The results showed that there is 8.77% increase in average power to the system with Power Management System compared to the MPPT controller alone.

## 1. 0 INTRODUCTION

The Philippines is an archipelagic country and some part of the population are located away from the electric grid. The introduction of Photovoltaic (PV) Panel makes it possible for these areas to have their own electricity and a possible solution for increasing demand in electricity. It offers clean and renewable energy, minimal maintenance and easy installation.

The graph shows the average irradiance (G) versus time from 8 AM to 6 PM here in the Philippines. Normally, the commercially available MPPT controllers neglect the power input after 5pm onwards.

Figure 2 shows that at different irradiances, there are different maximum power points. It can be seen that the voltage to produce maximum power also changes.

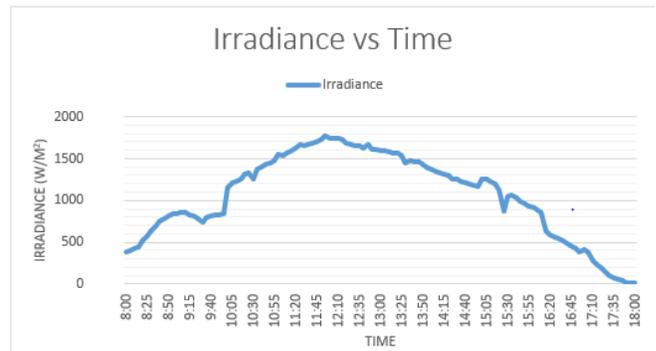

Fig.1. Average Irradiance vs. Time in the Philippines

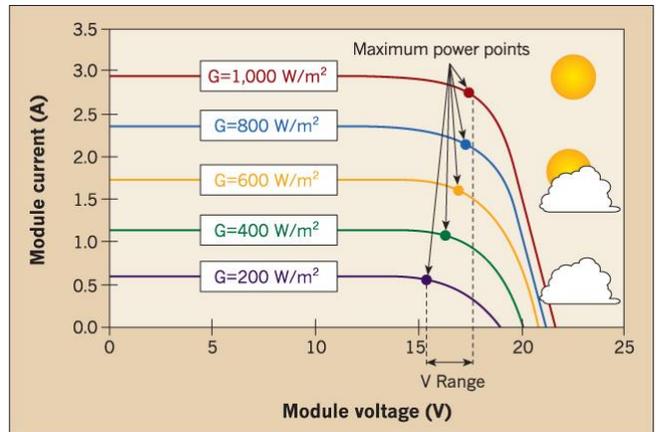

Fig.2. IV characteristic of a PV module (source: www.ecmweb.com)

Charge Controllers are used to maximize the energy harvested from PV Panels. Different techniques are studied to increase the efficiency of these controllers. One of those is Pulse Width Modulation. Pulse Width Modulation (PWM) controllers simply connect and disconnect the solar panel to the battery [1]. This controller detects the voltage between the PV panel and battery then if the voltage of the PV panel is greater than the charging voltage of the battery then it will connect the PV panel to the battery and vice versa. This technique is used to prevent the battery to be drained to the PV panel whenever the weather condition produces low irradiance. Another technique used to increase furthermore the efficiency of PWM controller is the Maximum Power Point Tracking (MPPT) Controller. It operates by changing duty cycle of DC-DC converter to attain the maximum power point [2-5]. By doing so, it increases the current that will flow to the battery, thus, reducing the overall system cost.

The commercially available MPPT controller combines the features of this technique, but it is limited to a certain voltage input and the charging voltage of the battery and this is easily affected by weather conditions.

The general objective of this study is to create a Power Management System that will increase the efficiency of a commercially available MPPT controller using Boost Converter and Switching circuit that will regulate the voltage of the PV panel to the required voltage of the MPPT controller to operate.

**2.0 REVIEW OF RELATED WORK**

*2.1 Design and Implementation of Maximum Power Point Tracking Solar Charge Controller*

The microcontroller that is used in the study is Arduino and showed an improved design of MPPT solar charge controller. The technique significantly reduced the systems loss, thus increasing the overall efficiency. The highest efficiency obtained in the study is 97.75% [6].

*2.2 Fuzzy Logic Controller of DC to DC DC-DC Converter for Energy Harvesting Applications*

The paper showed a design of a fuzzy logic controller that is used to drive the DC-DC converter for battery charging. It uses a constant voltage algorithm to maintain the output voltage of battery and serves as protection for overcharging and over voltage [7].

**3.0 METHODOLOGY**

*3.1 Block Diagram*

Figure 3 shows the overall block diagram of the system. A voltage sensor is attached to solar panel to detect its operating voltage and the microcontroller will decide if that voltage will go directly to the MPPT controller or to the Boost Converter first thru the Multiway Switch. In the case that the boost converter is activated, another voltage sensor is attached to the input of the MPPT controller to detect the output voltage of the boost converter so that the microcontroller could match the pulse width needed of the boost converter to attain the required voltage of the MPPT controller. The MPPT controller prioritizes the current requirement of load and the excess will go to the battery. It also has overvoltage protection for the load and battery.

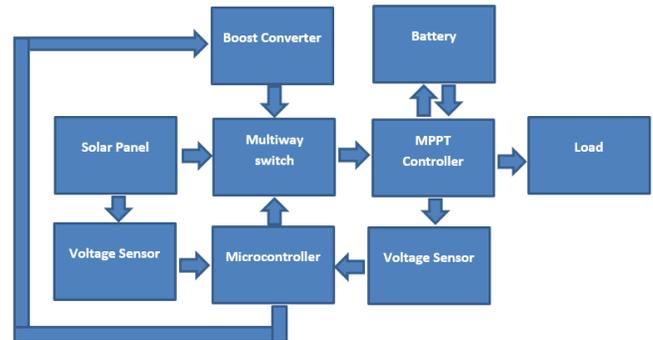

Fig.3. Block Diagram of the System

*3.2 Flowchart of the Power Management System*

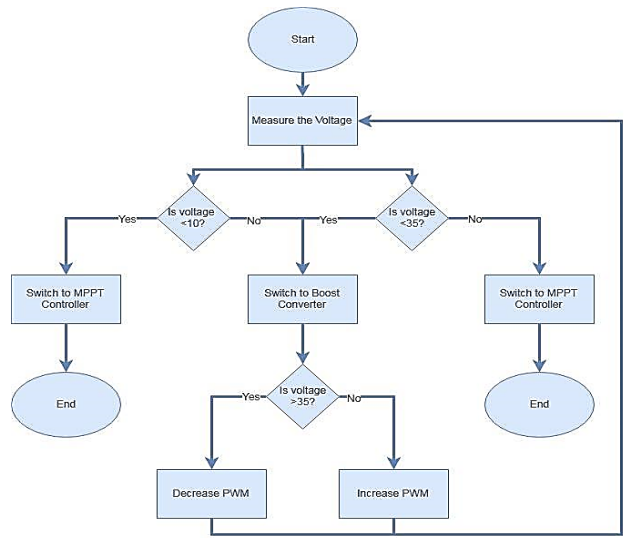

Fig.4. Flowchart of the Power Management System

Figure 4 shows the flowchart of the algorithm for the Power Management System. First, the voltage sensor will measure the voltage coming from the solar panel. If the voltage is greater than 10V and less than 35V, the relay will switch the path of the solar panel to the DC-DC converter then if the voltage is not equal to 35V the microcontroller will either increase or decrease the PWM to regulate the voltage to 35V. The DC-DC converter will stop working if it is lower than 10V because the voltage lower than that is harvested during sunset where the sun is not visible anymore.

*3.3   Schematic Diagram of the Power Management System*

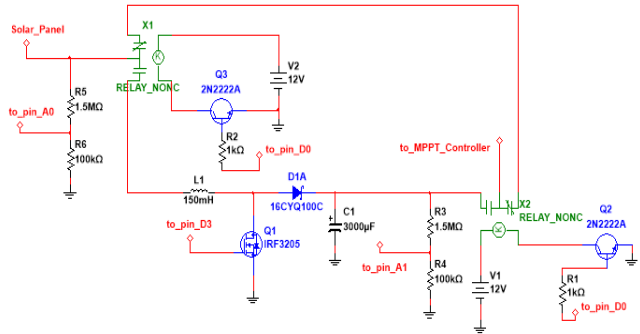

Fig.5. Schematic Diagram

Figure 5 shows the schematic diagram of the proposed Power Management System. The sensors are connected to the analog input of the Arduino namely A0 and A1. The Power MOSFET is connected to the PWM pin of Arduino namely D3. The relay drivers are controlled by pin D0. The frequency of PWM of Arduino is adjusted to 32.5kHz from 500Hz to make the circuit more efficient.

*3.3   Hardware*

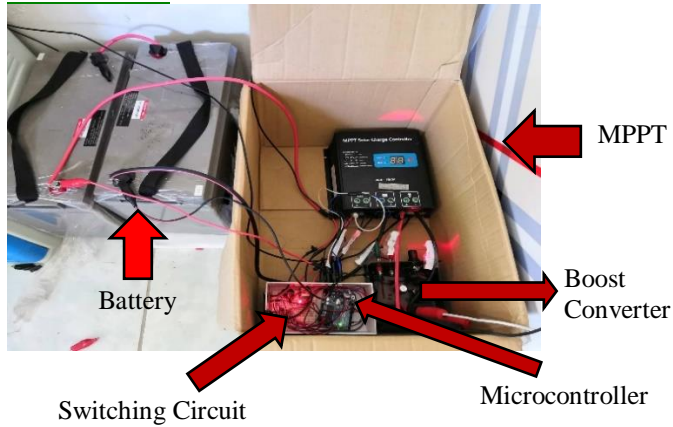

Fig.6. Actual Set Up

Figure 6 shows the actual set up of the circuit. The whole system is connected to a DC refrigerator as a load.

**4.0 RESULTS AND DISCUSSION**

*4.1   Testing of Power Management System*

Table 1. Verification of Power Management System

| Trial | Vin | Vout |
|---|---|---|
| 1 | 6 | 5.9 |
| 2 | 8 | 8 |
| 3 | 10 | 35 |
| 4 | 14 | 35 |
| 5 | 18 | 35.3 |
| 6 | 22 | 34.9 |
| 7 | 26 | 34.9 |
| 8 | 33 | 35.2 |
| 9 | 35 | 35 |
| 10 | 36 | 36.1 |
| 11 | 37 | 37 |
| 12 | 39.6 | 39.6 |
| 13 | 40 | 40 |

Table 1 shows that the Power Management System produces the desired output voltage according to the input voltage.

*4.2   Transfer Efficiency of Boost Converter*

Table 2. Transfer efficiency of the boost converter at different voltage inputs through a 100-ohm load

| Vin (V) | Power Input(W) | Power Output(W) | Transfer Efficiency (%) |
|---|---|---|---|
| 10 | 12.57 | 11.97 | 95.23 |
| 14 | 13.23 | 11.66 | 88.13 |
| 18 | 13.37 | 11.17 | 83.55 |
| 22 | 14.06 | 11.95 | 85 |
| 26 | 13.83 | 11.83 | 85.54 |
| 33 | 12.49 | 11.52 | 92.23 |
| 35 | 12.56 | 12.05 | 95.94 |

The average efficiency of the boost converter is 89.37% and based on the table 2 it ranges from 83.55% to 95.94%.

*4.3   Power Output of the system with and without the Power Management System*

Table 3. Average power output of the 2 system versus time

| TIME | MPPT only | With Power Management System |
|---|---|---|
| 8 AM | 42.35 W | 47.07 W |
| 9 AM | 66.11 W | 80.93 W |
| 10 AM | 107.98 W | 112.47 W |
| 11 AM | 134.23 W | 130.43 W |
| 12 NN | 137.47 W | 144.15 W |
| 1 PM | 122 W | 133.42 W |
| 2 PM | 100.93 W | 117.17 W |
| 3 PM | 87 W | 98.06 W |
| 4 PM | 49.07 W | 57.34 W |
| 5 PM | 10.87 W | 12.25 W |

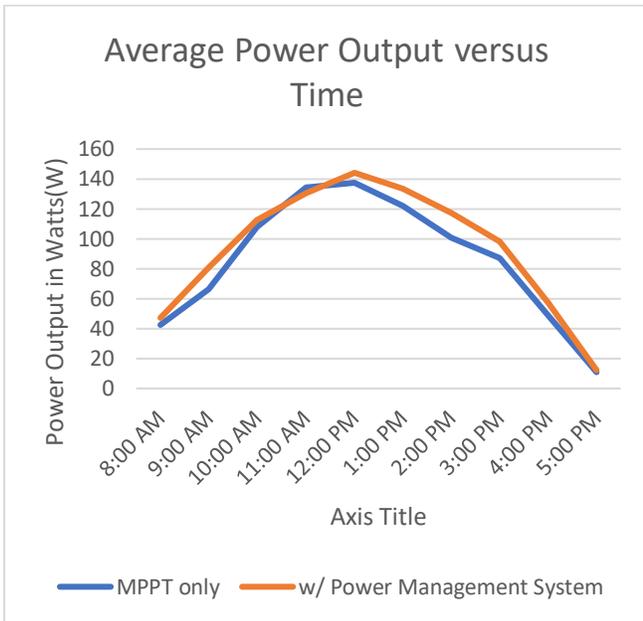

Fig.7. Graph of Average power output for the two systems

Table 3 and Figure 7 show the difference between the average power outputs produced by the 2 systems respect to time. It can be observed that the power produced by the system with Power Management System is higher in comparison with the system with MPPT Controller only. The average power for the system with Power Management System is about 93.33W which is 8.77% higher compared to 85.8W average power of the system with MPPT Controller only.

| T-test @ α = 0.05 | | |
|---|---|---|
| | Without Boost Converter | With Boost Converter |
| Mean | 85.85 | 99.1 |
| Observations | 121 | 121 |
| Pearson Correlation | 0.96 | - |
| t Stat | -2.31 | - |
| P(T<=t) one-tail | 0.01 | - |
| t Critical one-tail | 1.65 | - |
| P(T<=t) two-tail | 0.02 | - |
| t Critical two-tail | 1.97 | - |

Fig.8. Statistical Analysis

A t-test assuming unequal variances is conducted to check if the gain is significant. Figure 8 shows that t-value is less than the alpha value therefore the null hypothesis is rejected which states that there is no significant difference between the 2 systems. Therefore, the gain obtained earlier is significant.

## 5.0 CONCLUSION

The data and results show that the proponents were successful in designing and implementing the Power Management System that increases the efficiency of commercially available MPPT controllers. It prolongs the working hours of MPPT controllers which leads to a higher efficient system that can be used to places where the weather conditions are constantly varying.

## 6.0 RECOMMENDATIONS

The proponents recommend the use of microcontroller with higher frequency PWM function to further increase the transfer efficiency of the boost converter.

## 7.0 ACKNOWLEDGMENT

This study is supported by the University Research and Development Services Office and the University Research and Extension Council of the Technological University of the Philippines

### 9.0 ABOUT THE AUTHORS

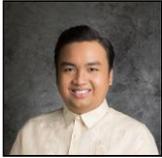

**Lean Karlo S. Tolentino** received his B.S. in electronics and communications engineering from the Technological University of the Philippines (TUP), Manila and M.S. in electronics engineering major in microelectronics from the Mapua University, in 2010 and 2015, respectively. He did his IC Design Faculty Immersion which was sponsored by the Department of Trade and Industry-Board of Investments (BOI) last 2017 at the National Sun Yat-sen University, Kaohsiung, Taiwan R.O.C. He is currently the Head of the Department of Electronics Engineering, TUP Manila. He is a member of IECEP, IEEE, ACM, and SPIE. His research interests include microelectronics and information and computing technologies.

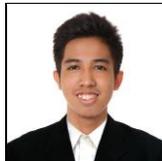

**Dan Frederico P. Abacco** received his B.S. in electronics engineering from the Technological University of the Philippines, Manila in 2019. He is a member of IECEP Manila Student Chapter. He passed the PhilNITS Information Technology Passport (IP) exam last 2018.

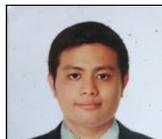

**Mar Jun M. Siquihod** received a diploma in electronics and communications engineering technology and B.S. in electronics engineering from the Technological University of the Philippines, Manila in 2016 and 2019, respectively. He is a member of IECEP Manila Student Chapter.